\documentclass[prl,twocolumn,floatfix,preprintnumbers,showpacs]{revtex4}
\usepackage{graphicx}
\usepackage{dcolumn}
\usepackage{bm}
\usepackage{color}

\providecommand{\adsurl}[1]{\href{#1}{ADS}}

\def\lsim{\mathrel{\mathop
  {\hbox{\lower0.5ex\hbox{$\sim$}\kern-0.8em\lower-0.7ex\hbox{$<$}}}}}
\def\gsim{\mathrel{\mathop
  {\hbox{\lower0.5ex\hbox{$\sim$}\kern-0.8em\lower-0.7ex\hbox{$>$}}}}}

\begin{document}
\newcommand{\mincir}{\raise
-2.truept\hbox{\rlap{\hbox{$\sim$}}\raise5.truept 
\hbox{$<$}\ }}
\newcommand{\magcir}{\raise
-2.truept\hbox{\rlap{\hbox{$\sim$}}\raise5.truept
\hbox{$>$}\ }}
\newcommand{\minmag}{\raise-2.truept\hbox{\rlap{\hbox{$<$}}\raise
6.truept\hbox
{$>$}\ }}

\newcommand{\half}{{1\over2}}
\newcommand{\bk}{{\bf k}}
\newcommand{\Ocdm}{\Omega_{\rm cdm}}
\newcommand{\ocdm}{\omega_{\rm cdm}}
\newcommand{\OM}{\Omega_{\rm M}}
\newcommand{\OB}{\Omega_{\rm B}}
\newcommand{\oB}{\omega_{\rm B}}
\newcommand{\OX}{\Omega_{\rm X}}
\newcommand{\cltt}{C_l^{\rm TT}}
\newcommand{\clte}{C_l^{\rm TE}}
\newcommand{\mwdm}{m_{\rm WDM}}
\newcommand{\mnu}{\sum m_{\rm \nu}}
\newcommand{\etal}{{\it et al.~}}
\newcommand{\lya}{{Lyman-$\alpha$~}}
\newcommand{\gad} {{\small {GADGET-2}}\,}
\input epsf

\title{Can sterile neutrinos be ruled out as warm dark matter candidates?}
\author{Matteo Viel,$^{1,2}$ 
Julien Lesgourgues,$^{3}$
Martin G.~Haehnelt,$^1$ 
Sabino Matarrese,$^{4,5}$ 
Antonio Riotto$^6$}
\affiliation{
$^1$Institute of Astronomy, Madingley Road, Cambridge CB3 0HA, United Kingdom\\
$^2$ INAF - Osservatorio Astronomico di Trieste, Via G.B. Tiepolo 11, 
I-34131 Trieste, Italy \\
$^3$Laboratoire d'Annecy-le-vieux de Physique Th\'eorique LAPTH, BP110,
F-74941 Annecy-le-vieux Cedex, France\\
$^4$ Dipartimento di Fisica ``G. Galilei'', Universit\`a di Padova,
Via Marzolo 8, I-35131 Padova, Italy\\
$^5$ INFN, Sezione di Padova,
Via Marzolo 8, I-35131 Padova, Italy\\
$^6$ CERN, Theory Division, CH 1211, Geneva 23, Switzerland
}
\date{\today}

\begin{abstract}
We present constraints on the mass of warm dark matter (WDM) particles
from a combined analysis of the matter power spectrum inferred from
the Sloan Digital Sky Survey \lya flux power spectrum at $2.2<z<4.2$,
cosmic microwave background data, and the galaxy power spectrum.  We
obtain a lower limit of $\mwdm \gsim 10$ ~keV ($2\sigma$) if the WDM
consists of sterile neutrinos and $\mwdm \gsim 2$~keV ($2\sigma$) for
early decoupled thermal relics. If we combine this bound with the
constraint derived from x-ray flux observations in the Coma cluster,
we find that the allowed sterile neutrino mass is $\sim 10$ keV (in
the standard production scenario). Adding constraints based on x-ray
fluxes from the Andromeda galaxy, we find that dark matter particles
cannot be sterile neutrinos, unless they are produced by a nonstandard
mechanism (resonant oscillations, coupling with the inflaton) or get
diluted by some large entropy release.
\end{abstract}


\maketitle
 
{\bf {Introduction.}}  Warm dark matter (WDM) has been advocated in
order to solve some apparent problems of standard cold dark matter
(CDM) scenarios at small scales (see \cite{bode} and references
therein), namely: the excess of galactic satellites, the cuspy and
high density of galactic cores, the large number of galaxies filling
voids. Moreover, recent observational results suggest that the shape
of the Milky Way halo is spherical \cite{Fellhauer:2006mt} and cannot
easily be reproduced in CDM models.  All these problems would be
alleviated if the dark matter (DM) is made of warm particles, whose
effect would be to suppress structures below the Mpc scale.  Detailed
studies of the dynamics of the Fornax dwarf spheroidal galaxy suggest
shallower cores than predicted by numerical simulations of CDM models
and put an upper limit on the mass of a putative WDM particle
\cite{fornax}. One of the most promising WDM candidate is a sterile
(right-handed) neutrino with a mass in the keV range, which could
explain the pulsar velocity kick \cite{pulsar}, help in reionizing the
universe at high redshift \cite{reio}, and emerge from many particle
physics models with grand unification
(e.g. \citep{sterileWDM,Abazajian:2005xn}). Because of a small,
non-zero mixing angle between active and sterile flavor states, X-ray
flux observations can constrain the abundance and decay rate of such
DM particles. The \lya absorption caused by neutral hydrogen
in the spectra of distant quasars is a powerful tool for constraining
the mass of a WDM particle since it probes the matter power spectrum
over a large range of redshifts down to small scales. In a previous
work, \cite{Viel:2005qj} used the LUQAS sample of high resolution
quasar absorption spectra to set a lower limit of 2 keV for the
sterile neutrino mass. More recently, exploiting the small statistical
errors and the large redshift range of the SDSS \lya forest data,
Seljak et al. \cite{Seljak:2006qw} found a lower limit of 14 keV.  If
the latter result is correct, a large fraction of the sterile neutrino
parameter space can be ruled out (assuming that all the DM is made of
sterile neutrinos); together with constraints from X-ray fluxes, this
discards the possibility that DM consists of sterile neutrinos
produced by non-resonant active-sterile neutrino oscillations
\citep{sterileWDM} (still, they could be produced by resonant
oscillations caused by a large leptonic asymmetry in the early
Universe \cite{Abazajian:2006yn}, or considerably diluted by some
large entropy release
\cite{Asaka:2006ek,Seljak:2006qw,Abazajian:2006yn}, or generated in a
radically different manner, e.g. from their coupling with the inflaton
\cite{Shaposhnikov:2006xi}). More recently, some joint analyses of the
SDSS flux power spectrum and the WMAP year three data \cite{wmap3}
have been presented in \cite{Viel:2006yh,Seljak:2006bg} for standard
$\Lambda$CDM models. The authors of \cite{Seljak:2006bg} found some
moderate disagreement between the inferred power spectrum
amplitudes. Instead, from an independent analysis of the SDSS data
\cite{Viel:2005ha}, the authors of \cite{Viel:2006yh} find good
agreement in their joint analysis.  Here, we extend the analysis of
\cite{Viel:2005ha} to constrain the mass of WDM particles.

{\bf {Data sets and Method.}} We use here the SDSS \lya forest data of
McDonald et al. \cite{McDonald:2004eu}, which consist of $3035$ quasar
spectra with low resolution ($R\sim 2000$) and low signal-to-noise
spanning a wide range of redshifts ($z=2.2-4.2$).  The data set
differs substantially from the LUQAS and C02 samples used
\cite{Viel:2005qj}, which contain mainly high resolution, high
signal-to-noise spectra at $z\sim2.5$.  More precisely, we use the 132
flux power spectrum measurements $P_F(k,z)$ that span 11 redshift bins
and 12 $k-$wavenumbers in the range $0.00141< k$ (s/km)$ <0.01778$
(roughly corresponding to scales of 5-50 comoving Mpc).  It is not
straightforward to model the flux power spectrum of the \lya forest
for given cosmological parameters, and accurate numerical simulations
are required.  McDonald et al. \cite{McDonald:2004eu} modelled the
flux power spectrum using a large number of Hydro Particle Mesh
simulations~\cite{HPM}, calibrated with a few small-box-size full
hydrodynamical simulations.  Here, instead, we model the flux power
spectrum using a Taylor expansion around a best fitting model: this
allows a reasonably accurate prediction of the flux power spectrum for
a large range of parameters, based on a moderate number of full
hydrodynamical simulations \cite{Springel:2005mi}.  The method has
been first introduced in Ref.\ \cite{Viel:2005ha} and we refer to this
work for further details.  The fiducial flux power spectrum has been
extracted from simulations of 60 $h^{-1}$ comoving Mpc and
$2\times400^3$ gas and DM particles (gravitational softening 2.5
$h^{-1}$ kpc) corrected for box size and resolution effects.  We
performed a number of additional hydrodynamical simulations with a box
size of 20 $h^{-1}$ comoving Mpc and with $2\times 256^3$ gas and DM
particles (grav. soft. 1 $h^{-1}$ kpc) for a WDM model with a
sterile neutrino of mass $m_s=1,4,6.5$ keV, to calculate the flux
power spectrum with respect to changes of the WDM particle mass. We
have checked the convergence of the flux power spectrum on the scales
of interests using additional simulations with $2\times 256^3$ gas and
DM particles and box sizes of 10 $h^{-1}$ Mpc (grav. soft. 0.5
$h^{-1}$ kpc).  We then used a modified version of the code CosmoMC
\cite{Lewis:2002ah} to derive the parameter likelihoods from the
combination of the \lya data with Cosmic Microwave Background
(CMB) and galaxy power spectrum data, from WMAP \cite{wmap3},
ACBAR~\cite{Kuo:2002ua}, CBI \cite{Readhead:2004gy},
VSA~\cite{Dickinson:2004yr} and 2dF \cite{Cole:2005sx}.  In total, we
used a set of 29 parameters: 7 cosmological parameters; 1 parameter
describing a free light-to-mass bias for the 2dF galaxy power
spectrum; 6 parameters describing the thermal state of the
Intergalactic Medium (parametrization of the gas temperature-gas
density relation $T=T_0(z)(1+\delta)^{\gamma(z)-1}$ as a broken power
law at $z=3$ in the two astrophysical parameters $T_0(z)$ and
$\gamma(z)$); 2 parameters describing the evolution of the effective
optical depth with redshift (slope and amplitude at $z=3$); 1
parameter which accounts for the contribution of damped \lya systems
and 12 parameters modelling the resolution and the noise properties
(see \cite{McDonald:2004xn}). We applied moderate priors to the
thermal history to mimic the observed thermal evolution as in
\cite{Viel:2004bf}, but the final results in terms of sterile neutrino
mass are not affected by this.

{\bf {Results.}} We assume the Universe to be flat, with
no tensor or neutrino mass contributions. We further note that adding CMB
and large scale structure data has very little effect on the results
for $m_{\rm s}$, since the free-streaming effect of WDM
particles is visible only on the scales probed by the \lya flux power
spectrum\footnote{in this work, the linear matter power spectrum
is computed under the assumption that the sterile neutrino phase-space
distribution is equal to that of active neutrinos multiplied by a
suppression factor \cite{CDW,Viel:2005qj}. Deviations from this first-order
approximation were computed in \cite{Abazajian:2005gj}, but typically these
corrections lower $m_s$ bounds by only 10\% \cite{Seljak:2006qw}.}.

\begin{figure}[h!]
\begin{center}
\includegraphics[angle=0,width=8.cm]{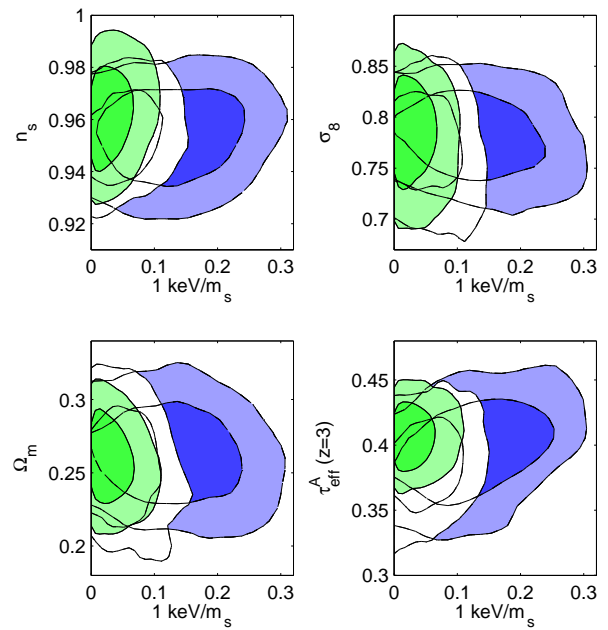}
\end{center}
\vspace{-0.5cm}
\caption{\label{f1} 2-dimensional marginalized likelihoods 
(68\% and 95\% confidence limits) for
$n_s,\sigma_8,\Omega_{\rm m}$ and the effective optical depth 
at $z=3$, using the SDSS data at $z\le 4.2$ (left, green),
$z\le 3.6$ (middle, white) and $z\le 3.2$ (right, blue).} 
\end {figure} 

In Figure \ref{f1} we show the 2-dimensional marginalized likelihoods
for the most important cosmological and astrophysical parameters:
$\sigma_8$, $n_s$, $\Omega_{\rm m}$ and the effective optical depth
amplitude measured at $z=3$, $\tau^A_{\rm eff}$, all plotted as a
function of the parameter $(1~\mathrm{keV})/m_{\rm s}$.  The
constraints on $m_{\rm s}$ get stronger for the \lya forest data in the
highest redshift bins. To demonstrate this we plot the likelihood
contours for data in three different redshift ranges: $z\le 3.2$ (blue), $z\le 3.6$ (white) and $z\le 4.2$ (green),
which is the whole data set.  The constraints improve by a factor
almost 3 (2) for the whole data set compared to the $z\le3.2$ ($z\le
3.6$) subsamples. At high redshifts, the mean flux level is lower and
the flux power spectrum is closer to the linear prediction making the
SDSS data points very sensitive to the free-streaming effect of WDM
\cite{Seljak:2006qw}. We find no strong degeneracies between $m_{\rm
s}$ and the other parameters, showing that the signature of a WDM
particle in the \lya flux power is very distinct, and that other
considered cosmological and astrophysical parameters cannot mimic its
effect.

\begin{figure}[h!]
\begin{center}
\hspace{0cm}
\includegraphics[angle=0,height=6.5cm,width=8cm]{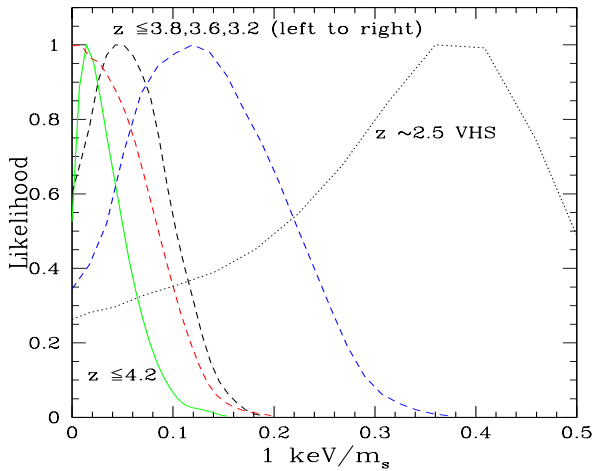}
\end{center}
\vspace{-0.5cm}
\caption{\label{f2} 1-dimensional marginalized likelihoods for the
parameter $(1~\mathrm{keV})/m_{\rm s}$ for the SDSS 
\lya data for the redshift ranges
$z\le 3.2,3.6,3.8,4.2$ and the VHS \citep{Viel:2004bf} data.}
\end{figure}

In Figure \ref{f2} we show the 1-dimensional marginalized likelihoods
for $(1~\mathrm{keV})/m_{\rm s}$ for several redshift ranges. The
$2\sigma$ lower limits for the sterile neutrino mass are: 3.9, 8.3,
8.1, 8.6, 10.3 keV for $z\le 3.2,3.4,3.6,3.8,4.2$, respectively.  The
corresponding limits for an early decoupled thermal relic are: 0.9,
1.7, 1.6, 1.7, 1.9 keV (see \cite{Viel:2005qj} for the correspondence
between the two cases). Also shown (dotted black line) is the
constraint obtained in \cite{Viel:2005qj} using the LUQAS and C02
samples \cite{Viel:2004bf,kimcroft}. The SDSS data improve
the constraint from the high resolution data at $z\sim 2.5$ by a factor
5.  This is mainly due to the extension to higher
redshift where the flux power spectrum is most sensitive to the effect
of WDM. The smaller statistical errors of the flux power spectrum and
the coverage of a substantial range in redshift help to break some of
the degeneracies between astrophysical and cosmological parameters and
also contribute to the improvement.  Our independent analysis confirms
the limits found in \cite{Seljak:2006qw} for the
SDSS \lya data and a small sample of high resolution data that also
extends to high redshift. Note, however, that our lower limit for
essentially the same data set is $\sim$30\% smaller (indeed, when using
only SDSS \lya data, Ref.\
\cite{Seljak:2006qw} obtains $m_{\rm s} >$ 12 keV ($2 \sigma$), which
includes a 10\% correction caused by the non-thermal momentum
distribution of sterile neutrinos \citep{Abazajian:2005gj}: so, for the
assumption made here, they would get $m_{\rm s} >$ 13 keV).

\begin{figure}[h!]

\begin{center}
\hspace{0cm}
\includegraphics[angle=0,width=9cm]{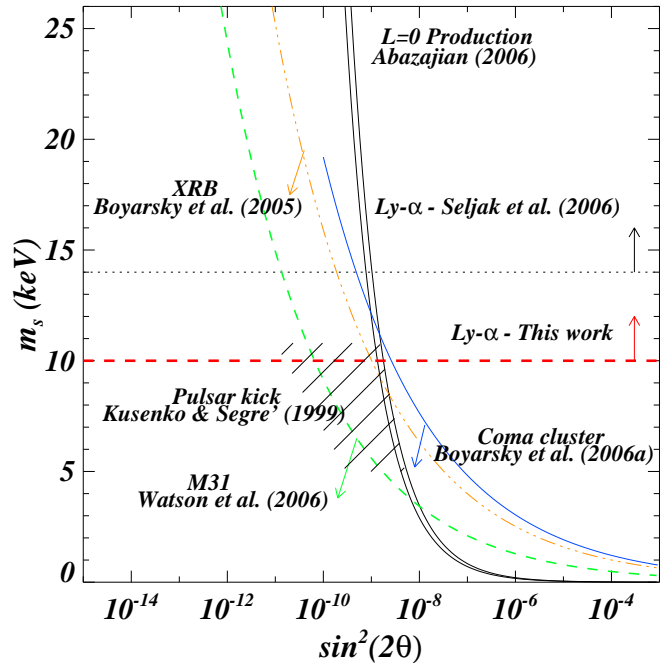}
\end{center}
\vspace{-0.5cm}
\caption{\label{f3} This plot summarizes some of the parameter
space constraints (at the 95\% C.L.)
for the sterile neutrino models, assuming that they
constitute the dark matter. Limits are explained in the text.}
\end{figure}

{\bf{Discussion.}}  In Figure \ref{f3} we summarize a number of
current constraints for sterile neutrinos in the $(m_s,\, \sin^2 \!
2\theta)$ plane, where $\theta$ is the vacuum $2\times2$ mixing angle
between active and sterile neutrinos \cite{Abazajian:2001vt}. We show
the limits obtained from different types of X-ray observations: X-ray
diffuse background (XRB, orange curve,
\cite{Boyarsky:2005us}); flux from the Coma cluster (blue curve,
\cite{Boyarsky:2006zi}); 
and finally, flux from the Andromeda galaxy (M31)
halo (95\% C.L., green dashed curve, \cite{Watson:2006qb}). 
In addition, we plot the \lya constraints obtained in this work (red
dashed) and in \cite{Seljak:2006qw} (black dotted). The region which
can explain observed pulsar kicks \cite{pulsar} is shown as the hatched
area.  Finally, according to \cite{Abazajian:2005xn}, sterile neutrinos
produced from non-resonant oscillations (i.e., in absence of
significant leptonic asymmetry, $L=0$) with a density $\Omega_{\rm DM}
= 0.23\pm0.04$ should lie between the two black solid curves (the
computation in \cite{Abazajian:2005xn} is based on simplifying
assumptions concerning the QCD phase transition; the effect of hadronic
corrections is currently under investigation \cite{Asaka:2006rw} and
could shift the allowed region in the $(m_s,\, \sin^2 \! 2\theta)$
plane).  If all these constraints are correct, then there is no room
for sterile neutrinos as DM candidates in the standard case.  Models in
which the decay of massive particles release some entropy and dilutes
the dark matter by a factor $S$ can alleviate the tension between the
\lya and X-ray bounds \citep{Asaka:2006ek}, but a very large $S$ is
needed \cite{Abazajian:2006yn,Seljak:2006qw}.  As mentioned in the
introduction, the sterile neutrino remains a viable WDM candidate for
alternative production mechanisms (e.g. resonant oscillations with $L
\neq 0$, or coupling with the inflaton). Recently, Ref.\
\cite{Abazajian:2006yn} questioned the results based on the LMC and Milky Way
because of uncertainties in modelling the dark matter distribution; and
also those based on detecting emission lines in cluster spectra
\cite{Boyarsky:2006zi}, which used a fixed phenomenological model for
X-ray emission (not shown in the figure but 30\% more constraining than
\cite{Boyarsky:2005us}). If these observational constraints are
inaccurate, then a sterile neutrino mass in the range $9\mincir m_s $
(keV) $\mincir 11.5$ and $\sin^2 \!  2\theta \sim2\times 10^{-9}$ would
be marginally consistent with the XRB bound
and the \lya forest data,
but it is strongly excluded by the robust limit 
obtained by Ref.~\cite{Watson:2006qb} (which is very conservative, since the
bound quoted as 2$\sigma$ by the authors requires 
a signal a few times larger than the background).
The corresponding emission line for such a
decaying sterile neutrino would be at $E\sim 5.5$ keV (close to, or
possibly contaminated by, the recently discovered Chromium line
\cite{Werner:2005rf}).  If instead all X-ray constraints are correct,
but the two recent \lya forest constraints are not accurate, then a
mass of $m_s\sim 2$~keV is still possible and compatible with the
robust and conservative lower limit from \cite{Viel:2005qj}.  It would
also satisfy the requirement from the dynamical analysis of the Fornax
dwarf galaxies \cite{fornax}. However, the latter possibility appears
unlikely. Even if the highest redshift bins of the SDSS \lya forest
data were affected by not yet considered systematic errors the analysis
of the data with $z\le 3.2$ still gives a lower limit of about $\sim$
3.5 keV (see \cite{Watson:2006qb}). Appealing to an insufficient
resolution of the hydrodynamical simulations would also not help, since
an increase in resolution could only increase the flux power spectrum
at small scales and raise the lower limits. We have furthermore checked
explicitly that this is not the case and that other possible effects on
the flux power have a different signature than that of WDM.  A
potentially big improvement on the quality of the constraints from \lya
forest data could be achieved by an analysis of a large set of
high-redshift, high-resolution data to extend the measurement of the
flux power spectrum at high redshift to smaller scales.  This would,
however, also require accurate modelling of the thermal history and the
contribution of associated metal absorption to the small scale flux
power spectrum.

{\bf {Acknowledgements.}}  Simulations were done at the UK Cosmology
Supercomputer Center in Cambridge funded by PPARC, HEFCE and Silicon
Graphics/Cray Research. We thank A. Lewis for technical help and
K. Abazajian, S. Hansen, A. Kusenko, J. Sanders, M. Shaposhnikov and
C. Watson for useful comments.

\end{document}